\documentclass{PoS}

\title{Observational Properties of Feebly Coupled Dark Matter}

\ShortTitle{Observational Properties of Feebly Coupled Dark Matter}

\author{{Matti Heikinheimo}%
        \\
       Helsinki Institute of Physics and University of Helsinki\\
       E-mail: \email{matti.heikinheimo@helsinki.fi}}
       
\author{\speaker{Tommi Tenkanen}%
        \\
       Helsinki Institute of Physics and University of Helsinki\\
       E-mail: \email{tommi.tenkanen@helsinki.fi}}
       
       \author{{Kimmo Tuominen}%
        \\
       Helsinki Institute of Physics and University of Helsinki\\
       E-mail: \email{kimmo.i.tuominen@helsinki.fi}}
       
       \author{{Ville Vaskonen}%
        \\
       Helsinki Institute of Physics and University of Jyv\"askyl\"a\\
       E-mail: \email{ville.vaskonen@jyu.fi}}

\abstract{We show that decoupled hidden sectors can have observational consequences. As a representative model example, we study dark matter production in the Higgs portal model with one real singlet scalar $s$ coupled to the Standard Model Higgs via $\lambda_{\rm hs}\Phi^\dagger\Phi s^2$ and demonstrate how the combination of non-observation of cosmological isocurvature perturbations and astrophysical limits on dark matter self-interactions imply stringent bounds on the magnitude of the scalar self-coupling $\lambda_{\rm s}s^4$. For example, for dark matter mass $m_{\rm s}=10$ MeV and Hubble scale during cosmic inflation $H_*=10^{12}$ GeV, we find $10^{-4}\lesssim \lambda_{\rm s}\lesssim 0.2$.
}

\FullConference{38th International Conference on High Energy Physics\\
          3-10 August 2016\\
         Chicago, USA}

\begin{document}

In light of the current experimental bounds on dark matter (DM) properties \cite{Klasen:2015uma}, the elusive nature of DM raises an important question: if dark matter interacts so feebly with the Standard Model (SM) particles that it can never be detected by direct experiments, is all hope lost in probing the properties of DM beyond its gravitational interactions? In this proceeding, based on \cite{Heikinheimo:2016yds}, we discuss how the future observations of galaxy cluster dynamics and the cosmic microwave background (CMB) may be used to probe DM properties even if the current or near-future direct or indirect detection experiments will see nothing.

We concentrate on a model where DM resides in a hidden sector, which contains only a real singlet scalar field $s$ coupled to the SM sector via the Higgs portal $\lambda_{\rm hs} \Phi^\dagger\Phi s^2$, where $\sqrt{2}\Phi=(0,h+v)$ is the SM Higgs doublet. Here $v=246$ GeV is the vacuum expectation value (vev) of the Higgs field. For simplicity, we impose a $Z_2$ symmetry for $s$ to guarantee its stability on cosmological time scales, and thus its viability as a DM candidate. The scalar potential of the model is therefore given by
\begin{equation}
V = \mu_{\rm h}^2\Phi^\dagger\Phi + \lambda_{\rm h}(\Phi^\dagger\Phi)^2  + \frac{\mu_{\rm s}^2}{2} s^2 + \frac{\lambda_{\rm s}}{4}s^4 + \frac{\lambda_{\rm hs}}{2} \Phi^\dagger\Phi s^2 ,
\end{equation}
where we assume $\mu_{\rm s}^2,\lambda_{\rm s}>0$. After the electroweak symmetry breaking at $T_{\rm EW}\simeq 100$ GeV the physical masses are given by $m_{\rm s}^2=\mu_{\rm s}^2+\lambda_{\rm hs}v^2/2$ and $m_{\rm h}^2=\mu_{\rm h}^2+3\lambda_{\rm h}v^2$.
 
If the coupling $\lambda_{\rm hs}$ between the SM sector and the hidden sector is very small, $|\lambda_{\rm hs}|\lesssim 10^{-7}$, the sectors were never in thermal equilibrium with each other in the early universe. In that case, the observed DM abundance is produced by a freeze-in mechanism \cite{McDonald:2001vt, Hall:2009bx} dominated by Higgs boson decays $h\to ss$, assuming $m_{\rm h}\geq 2m_{\rm s}$.
The yield begins when the Higgs field acquires a vev and starts to decay into $s$ particles, and stops when the universe cools down so that the annihilations of SM particles cannot produce decaying Higgs particles any more. In the absence of large DM self-interactions, the final DM abundance is
\begin{equation}
\label{FI_abundance}
\frac{\Omega_{\rm DM}h^2}{0.12} = 5\times10^{19}\lambda_{\rm hs}^2\left(\frac{m_{\rm s}}{10\,{\rm MeV}}\right),
\end{equation}
implying a small value of the portal coupling, compatible with the key assumption of the freeze-in scenario that the dark matter particles were not thermalized with the SM particles above $T_{\rm EW}$.

However, depending on the DM self-interaction strength, the hidden sector may thermalize within itself. This happens if the magnitude of $\lambda_{\rm s}$ exceeds a critical value $\lambda_{\rm s}^{\rm crit.}=\lambda_{\rm s}^{\rm crit.}(\lambda_{\rm hs}, m_{\rm s})$, whose analytical derivation and the resulting DM abundance in different cases can be found in \cite{Heikinheimo:2016yds}. If the hidden sector thermalizes, the final DM abundance is determined by freeze-out occurring in the hidden sector \cite{Carlson:1992fn}. 

The presence of relatively large DM self-interactions is favored by observations of flat cores of galactic DM halos \cite{Moore:1994yx,Flores:1994gz} and spatial offset of galactic DM halo from stars in Abell 3827 galaxy cluster \cite{Massey:2015dkw}. Also cosmological constraints on decoupled hidden sector scalars favor non-zero scalar self-interactions. As originally shown in \cite{Kainulainen:2016vzv}, the non-observation of isocurvature perturbations in the CMB places a constraint
\begin{equation}
\label{cosmolimit}
\lambda_{\rm s} \gtrsim 4\times 10^{-8}\left(\frac{m_{\rm s}}{10\,{\rm MeV}}\right)^{8/3}\left(\frac{H_*}{10^{11}\,{\rm GeV}}\right)^4 ,
\end{equation}
where $H_*$ is the Hubble scale during cosmic inflation. The limit (\ref{cosmolimit}) applies for any scenario where the singlet scalar $s$ is a light and energetically subdominant spectator field during inflation with a usual perturbation spectrum, and where the final DM abundance is determined by the standard freeze-in mechanism. Cosmological observations therefore provide a lower bound on scalar DM self-interactions. 

\begin{figure}
\begin{center}
\includegraphics[width=.65\textwidth]{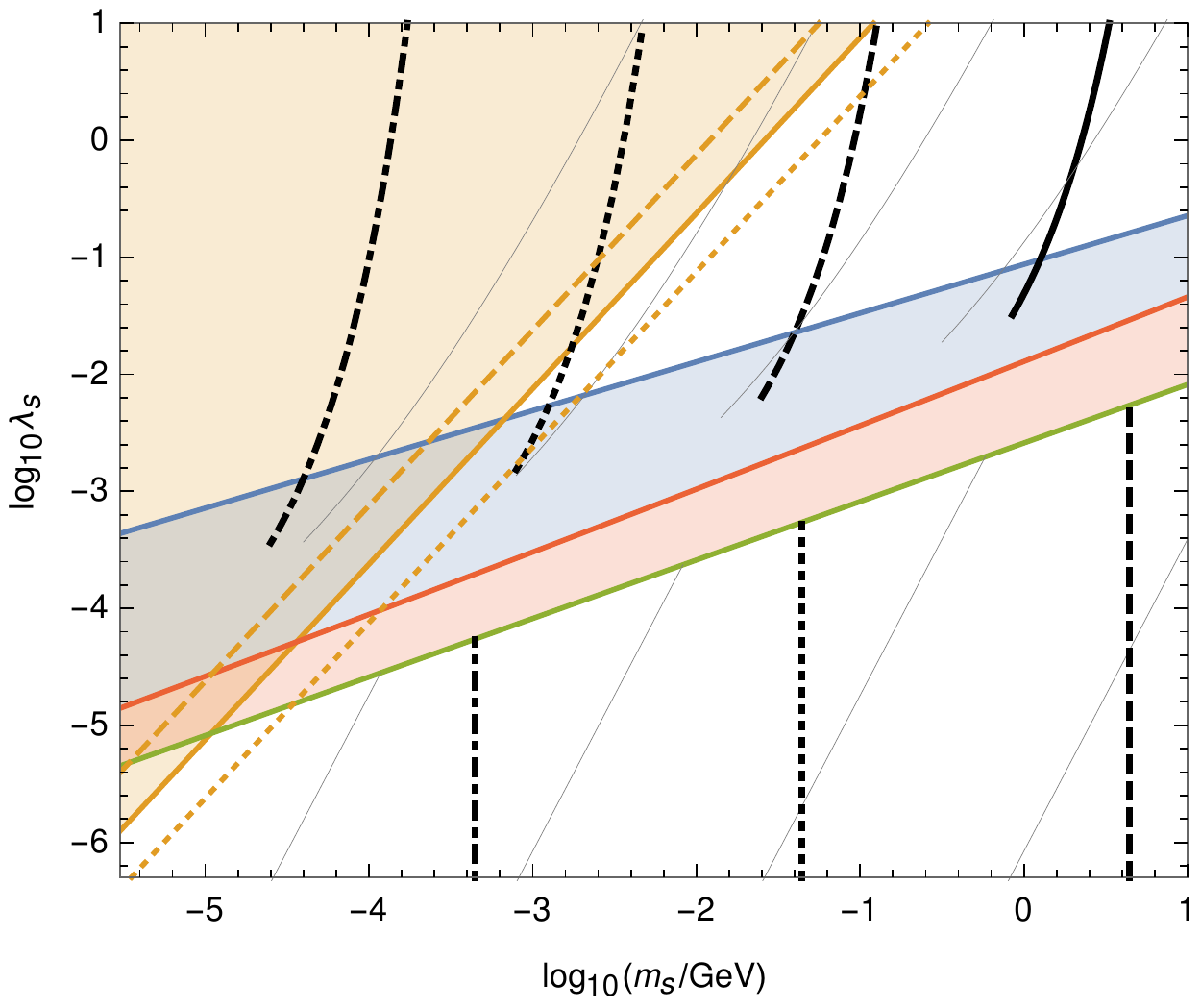}
\caption{The dark matter self-coupling as a function of dark matter mass required to obtain the correct relic abundance for $\lambda_{\rm hs}=10^{-9},10^{-10},10^{-11},10^{-12}$ (black lines, from left to right), superimposed over the DM self-interaction bound and isocurvature constraints. See text for the blue, red, and white regions exhibiting different dynamical origin for the DM abundance. The self-interaction limit, $\sigma_{\rm DM}/m_{\rm DM} < 1$ cm$^2$/g, is shown by the yellow shaded region in the top left corner together with the $\sigma_{\rm DM}/m_{\rm DM} = 10, 0.1$ cm$^2$/g contours (dashed and dotted, respectively), and the isocurvature constraints by the gray contours for \mbox{$H_* = 10^{13}, 10^{12}, 10^{11}, 10^{10}$ GeV} from left to right. The area to the right of each contour is ruled out for the given value of the inflationary scale $H_*$. The Figure is from Ref. \cite{Heikinheimo:2016yds}.}
\label{DM_constraints}
\end{center}
\end{figure}

On the other hand, observations of galaxy cluster mergers (such as the Bullet Cluster) provide an upper bound on dark matter self-interactions, $\sigma_{\rm DM}/m_{\rm DM}\lesssim \mathcal{O}(1)\, {\rm cm}^2$/g \cite{Markevitch:2003at}. In the limit $m_{\rm s}\ll m_{\rm h}$ the scalar self-interaction cross-section divided by the scalar mass is $\sigma_{\rm s}/m_{\rm s} \simeq 9\lambda_{\rm s}^2/(32 \pi m_{\rm s}^3)$, so that together the cosmological and astrophysical constraints imply 
\begin{equation}
\label{scalar_ia_bound}
4\times 10^{-8}\left(\frac{m_{\rm s}}{10\,{\rm MeV}}\right)^{8/3}\left(\frac{H_*}{10^{11}\,{\rm GeV}}\right)^4 \lesssim \lambda_{\rm s} \lesssim 0.2\left(\frac{m_{\rm s}}{10\,{\rm MeV}}\right)^{3/2} .
\end{equation}
These limits are depicted in Figure \ref{DM_constraints} together with the resulting DM abundance.

Above the blue shaded region the DM abundance is produced by the dark freeze-out mechanism at a temperature where DM is non-relativistic. Within the blue shaded region the dark freeze-out happens at a (semi)relativistic temperature $3T_{\rm FO}\geq m_{\rm s}$, and in the lower white region the DM abundance can be obtained via the standard freeze-in scenario. In the red region the analytical method used in \cite{Heikinheimo:2016yds} yields no solutions. The allowed parameter space is limited by the constraints from Lyman-$\alpha$ forest data, excluding warm DM with mass below $m_{\rm s} \leq 3$ keV \cite{Viel:2013apy}. 

If the hidden sector thermalizes with itself and the resulting DM abundance is determined by dark freeze-out rather than the standard freeze-in, the isocurvature bound on DM self-interactions differs from the simple result (\ref{cosmolimit}) and has to be calculated by following the hidden sector dynamics more carefully \cite{Heikinheimo:2016yds}. The effect of this correction is to increase the importance of the isocurvature constraints, which is seen in the shift of gray contours above the red shaded region in Figure \ref{DM_constraints}.

The results demonstrate how valuable information about DM models can be extracted from combinations of cosmological and astrophysical observables even in the case where new physics is beyond the reach of direct or indirect DM detection experiments. Finally, we emphasize that even though we have concentrated on a simple Higgs portal scenario with one real singlet scalar field only, qualitatively similar results are expected to constrain the mass and coupling values also in other, more generic portal model -type extensions of the SM.

\section*{Acknowledgements}
This work is financially supported by the Academy of Finland, grant \#267842. TT acknowledges funding from the Research Foundation of the University of Helsinki and the Emil Aaltonen foundation. VV acknowledges funding from the Magnus Ehrnrooth foundation.

\end{document}